\documentclass[aps,showpacs,twocolumn,prb,superscriptaddress]{revtex4}
\usepackage{amsfonts}
\usepackage{amsmath}
\usepackage{amssymb}
\usepackage{wasysym}
\usepackage{graphicx}

\newcommand{\placefigure}[4]
{
 \begin{figure}[t]
 \includegraphics[width=#2]{#1}
 \caption{#3}
 \label{#4}
 \end{figure}
}

\newcommand{\br}{{\bf{r}}}

\begin{document}
\preprint{}

\title{Relation between directed polymers in random media and
  random bond dimer models}

\author{Ying Jiang}
\affiliation{Department de Physique, Universit\'e de Fribourg,
Chemin du Mus\'ee 3, CH-1700 Fribourg, Switzerland}

\author{Thorsten Emig}
\affiliation{Laboratoire de Physique
  Th\'eorique et Mod\`eles Statistiques, CNRS UMR 8626, Universit\'e
  Paris-Sud, 91405 Orsay, France}

\keywords{} \pacs{05.20.-y, 75.10.Nr, 05.50.+q}

\begin{abstract}
  We reassess the relation between classical lattice dimer models and
  the continuum elastic description of a lattice of fluctuating
  polymers. In the absence of randomness we determine the density and
  line tension of the polymers in terms of the bond weights of
  hard-core dimers on the square and the hexagonal lattice. For the
  latter, we demonstrate the equivalence of the canonical ensemble for
  the dimer model and the grand-canonical description for polymers by
  performing explicitly the continuum limit. Using this equivalence
  for the random bond dimer model on a square lattice, we resolve a
  previously observed discrepancy between numerical results for the
  random dimer model and a replica approach for polymers in random media.
Further potential applications of the equivalence are briefly discussed.
\end{abstract}

\maketitle

%%%
%%% Introduction
%%%
\section{Introduction}

Dimer coverings of different lattice types have been employed recently
as a starting point to study more complex physical systems like quantum
dimer models \cite{Rokhsar-Kivelson-PRL,Moessner-RVB}, geometrically
frustrated Ising magnets with simple quantum dynamics induced by a
transverse magnetic field \cite{Moessner-Sondhi-PRB,Jiang-Emig-PRL}
and elastic strings pinned by quenched disorder
\cite{Zeng-Middleton-Shapir-PRL,Bogner-Emig-Taha-Zeng-PRB}. The common
concept of these approaches is to add to a classical dimer model with a
hard-core interaction a perturbation in form of simple quantum
dynamics, quenched disorder (random bonds) or additional (classical)
dimer interactions. The classical hard-core dimer model can be solved
on arbitrary planar graphs
\cite{Kasteleyn-61-Fisher-61,Kasteleyn-63-Fisher-66}. For bipartite
lattices there exists a representation of the dimer model in terms of
a height profile of a two-dimensional surface
\cite{Henley-JSP,Bloete}. Steps separating terraces of equal height
form a lattice of directed and non-crossing polymers
\cite{Zeng-Middleton-Shapir-PRL}. This polymer lattice has been used
to study the effect of quantum fluctuations for a Ising
antiferromagnet on a triangular lattice in a transverse field
\cite{Moessner-Sondhi-PRB,Jiang-Emig-PRL,Jiang-Emig-preprint-2005}. Moreover,
the dimer model with random bond energies can be simulated in
polynomial time and provides an independent test of the replica
theory for the pinning of elastic polymers by quenched disorder
\cite{Nattermann-Scheidl-AP}. Hence it is important to understand the
relation between lattice dimer models and the elastic continuum
description of the corresponding polymers and to relate the parameters
of the two models.

Here we show that the canonical dimer ensemble in the classical case
maps to the grand canonical ensemble of the polymer system. Regarding
the polymers as imaginary-time world lines of free fermions in one
dimension, we give a simple derivation of the continuum free energy of
lattice dimer models which agrees with the exact result if the
continuum limit is taken properly. Applying the relation between
dimers and polymers to the dimer model with random bond energies, we
resolve a previously observed discrepancy between numerical
simulations of the dimer system and a replica theory for the polymers.
We analyze which quantities of the pinned polymers can be probed by
simulations of the random dimer model. More specifically, we study the
dimer model both on the hexagonal and the square lattice, and discuss
the meaning of the different lattice symmetries for the polymer
representation.  In the presence of bond disorder, we focus on the
square lattice since its polymer density is conserved independently of
the disorder configuration, and hence shows no sample-to-sample
variations. Our results should provide a starting point for other
situations where no exact solution of the dimer model is possible as,
e.g., in a recently studied case of nearest neighbor dimer interactions
\cite{Alet+05}.  The analogy between directed polymers in two
dimensions and Luttinger liquids could be applied to understand more
general interacting dimer models.

%%%
%%% Models (Dimer & elastic) and its relation
%%%

\section{Models: Dimers and polymers}

The partition function of the dimer model on the hexagonal lattice is
given by
\begin{equation}
\label{eq:Z-hexa}
Z_{\hexagon}= \sum_{\{D\}} z_1^{n_1} z_2^{n_2}\, ,
\end{equation}
where the sum runs over all complete hard-core dimer coverings of the
lattice, and $n_1$ and $n_2$ are the numbers of dimers occupying the
two types of non-vertical bonds of the hexagonal lattice, see
Fig.\ref{fig:mapping}, which carry weights $z_1$ and $z_2$, respectively.
The weights on the vertical bonds are assumed to be unity.  Using the
same notation, we define for the square lattice the partition function
as
\begin{equation}
\label{eq:Z-squa}
Z_{\Box}= \sum_{\{D\}} z_1^{n_1} \, ,
\end{equation}
where $n_1$ is now the number of dimers covering horizontal bonds
which have a weight of $z_1$ while all vertical bonds carry unity as
weight. For these clean dimer models the
partition function and correlation functions are known from exact
results \cite{Kasteleyn-63-Fisher-66,Kasteleyn-61-Fisher-61}.  For
random weights, only numerical results are available, see, e.g.,
Refs.~\onlinecite{Zeng-Middleton-Shapir-PRL,Bogner-Emig-Taha-Zeng-PRB}.
However, there is a useful connection between the dimer models and
non-crossing directed polymers in $(1+1)$ dimensions. This relation is
independent of the actual bond energies, and hence applies also to
{\it random} dimer models. This is particularly interesting since the
random bond energies translate to a pinning potential for the
polymers, a problem whose continuum version can be studied in $(1+1)$
dimensions by a replica Bethe ansatz
\cite{Kardar-NPB,Emig-Kardar-NPB}. Numerical algorithms for the
random dimer model provide hence a unique opportunity to probe the
replica symmetric theory which is commonly used to describe pinning of
elastic media.

The relation between dimers and polymers is established by superposing
every dimer configuration by a fixed reference dimer
configuration. For the hexagonal lattice, the reference state consists
of a covering of all vertical bonds, whereas for the square lattice a
staggered covering of the vertical bonds is chosen, see
Fig.~\ref{fig:mapping}. In the superposition state, a bond is covered
by a dimer (or polymer segment) if either it is covered only in the
original state or only in the reference state. Because of the hard
core constraints for the dimers, the polymers are non-crossing and
they are oriented along the vertical direction due to the choice of the
reference state.

The resulting lattice of polymers can be described in the
{\it continuum} limit by an elastic theory which is of the form
\begin{equation}
\label{cont_lines}
H_{\rm el}=\int\! d^2\!\br \left\{\frac{c_{11}}{2}(\partial_x u)^2 +
\frac{c_{44}}{2}(\partial_y u)^2
+\rho(\br)V(\br)\right\}
\end{equation}
with compression modulus $c_{11}$, tilt modulus $c_{44}$ and local
polymer density $\rho(\br)=\sum_j \delta(x-x_j(y))$, where $x_j(y)$ is
the path of the $j^{\rm th}$ polymer. The random bond energies are
accounted for by a random pinning potential $V(\br)$ which is
uncorrelated, i.e.,
\begin{equation}
\label{eq:VV-corr}
\overline{V(\br)V(\br')}=\Delta\delta(\br-\br')
\end{equation}
so that $\Delta$ measures the strength of disorder. In order to
compare results for the dimer and the polymer model, we
establish a relation between the dimer weights and the elastic
constants and mean density of the polymers. Let us consider first the
clean limit with $\Delta\equiv 0$. It is obvious from the mapping
between dimers and polymers that the polymer density can vary with the
dimer covering. For example, if the dimer state matches exactly the
reference state, the polymer density is zero. However, one can define
a {\it mean} polymer density by averaging over all dimer coverings.

For the hexagonal lattice the mean density is determined by the mean
number of occupied non-vertical bonds in the original dimer
configuration so that $\rho_{\hexagon}= \langle n_1 + n_2
\rangle/(\sqrt{3}b_{\hexagon} N)$ where $N$ is the total number of
dimers, $b_{\hexagon}$ the lattice constant and $\langle \ldots
\rangle$ denotes here an ensemble average over all dimer
coverings. This yields
\cite{Jiang-Emig-preprint-2005,Yokoi-Nagle-Salinas-JSP}
\begin{equation}
\label{eq:hexa-density}
\rho_{\hexagon}=\frac{2}{\pi\sqrt{3}b_{\hexagon}}
\arcsin\left[\frac{\left(z_{1}+z_{2}\right)^{2}-1}{4z_{1}z_{2}}\right]^{1/2}
\end{equation}
if $z_1+z_2>1$ and $\rho_{\hexagon}=0$ if $z_{1}+z_{2}\leq 1$.  For
the square lattice, the mean number of polymers is determined by the
probability that a vertical bond is occupied by a segment of the
polymer. This is case if the bond {\it is covered} by a dimer in the
original dimer configuration and {\it not covered} in the reference
state, or in the other way around. The probability that a vertical
bond is covered by a dimer in the original covering is $p_d = 1/2 -
\phi(z_1)$ with $\phi(z_1)=\arctan(z_1)/\pi$. For the reference state
it is simply $p_r=1/2$. Hence after the superposition of the two dimer
states, the probability that a vertical bond is covered by a polymer
is given by $p_d(1-p_r)+p_r(1-p_d)=1/2$ independent of $z_1$. This fixes
the mean density at \cite{Bogner-Emig-Taha-Zeng-PRB}
\begin{equation}
\label{eq:square-density}
\rho_\Box=\frac{1}{2b_\Box} \, ,
\end{equation}
where $b_\Box$ is the lattice constant.  Notice that the density on
the square lattice does not change with the bond weight $z_1$.

The elastic constants are length scale dependent due to
renormalization effects from the non-crossing constraint.  Since their
is no additional interaction between the dimers than the hard core
repulsion, the compression modulus $c_{11}$ is zero on microscopic
scales. A finite macroscopic $c_{11}$ is generated by a reduction of
entropy due to polymer collisions, see below. The tilt modulus
$c_{44}=g\rho$ on microscopic scales (or at very low density) is given
by the line tension $g$ of a single polymer and the mean polymer
density.  The reduced line tension $g/T$ of an individual polymer at
temperature $T$ can be obtained from a simple random walk on the
lattice \cite{Jiang-Emig-preprint-2005} which performs transverse
steps according to the weights of the dimer model.  For the hexagonal
lattice it reads \cite{Jiang-Emig-preprint-2005}
\begin{equation}
\label{eq:hexa-g}
\frac{g_{\hexagon}}{T}=\frac{2+\eta+1/\eta}{2b_{\hexagon}}
\end{equation}
with $\eta=z_{1}/z_{2}$, and for the
square lattice one has
\begin{equation}
\label{eq:square-g}
\frac{g_\Box}{T}=\frac{2z_1+1}{2z_1 b_\Box}.
\end{equation}
From this result we see that the polymers become stiffer if one
decreases the weights on (one type of) the non-vertical bonds, hence
preventing transverse wandering.

\placefigure{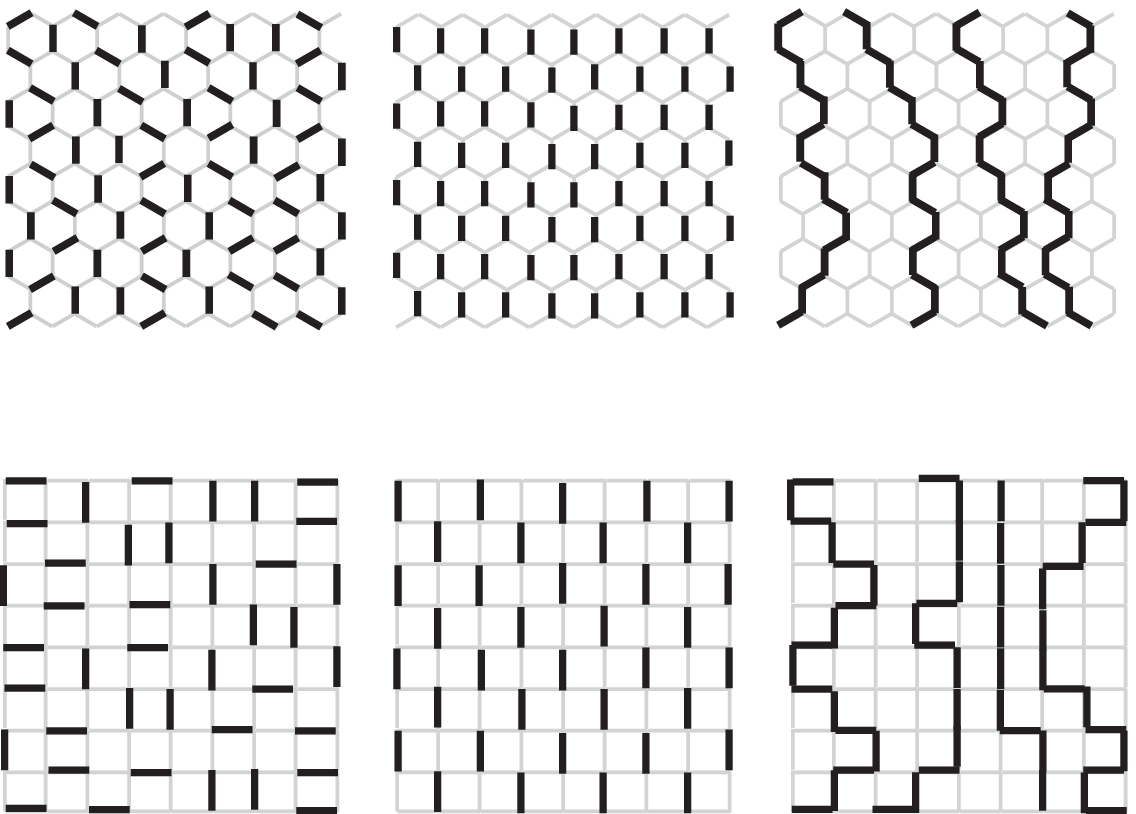}{0.9\linewidth}{Mapping of dimer
  configurations to directed polymers by superposition with a fixed
  reference covering (middle).}{fig:mapping}

%%%
%%% clean case
%%%
\section{Clean system: continuum limit and thermodynamic
ensembles}

% hexagonal lattice

Before treating the random system, let us first compare the free
energy density of the dimer models and the corresponding polymer
lattice by taking the continuum limit. The free energy of dimer model
on the hexagonal lattice can be computed exactly
\cite{Yokoi-Nagle-Salinas-JSP}. By changing variables from $z_1$,
$z_2$ to $\eta=z_1/z_2$ and $\rho_{\hexagon}$, one can express the
free energy density in terms of the physical quantities of the
polymers on the lattice \cite{Jiang-Emig-preprint-2005}.  The
result is
\begin{equation}
\label{eq:f-hexa-exact}
f_{\rm dimer} =
-  \frac{2\pi}{\sqrt{3}} \int_0^{\rho_{\hexagon}} d\rho'
\frac{\eta \rho' \sin (\pi \sqrt{3} b_{\hexagon} \rho')}
{1+ \eta^2 + 2\eta \cos(\pi \sqrt{3} b_{\hexagon} \rho') } \, ,
\end{equation}
where the energy is measured relative to the line $z_1+z_2=1$ of
vanishing polymer density.  For this expression we can take explicitly
the continuum limit by sending $b_{\hexagon} \to 0$ while keeping the
polymer density $\rho_{\hexagon}$ fixed, i.e., we adjust $z_1$ and
$z_2$ so that the arcsin in Eq. \eqref{eq:hexa-density} tends to zero
$\sim b_{\hexagon}$ while $\eta=z_1/z_2$ is kept fixed. This yields
the free energy density for the continuum version of the dimer model,
expressed in terms of the polymer parameters,
\begin{equation}
f_{\rm dimer}=-\frac{\pi^{2}}{3}\frac{T}{g_{\hexagon}}\rho_{\hexagon}^{3} \, ,
\label{dimer-free-energy}
\end{equation}
where we used Eq.~\eqref{eq:hexa-g}.

An alternative approach to treat a system of interacting polymers in
$(1+1)$ dimensions is to regard each polymer as a world line of a
boson in imaginary time, leading to the quantum theory of interacting
bosons in one spatial dimension.  Choosing the imaginary time
direction along the occupied bonds in the dimer reference state, the
non-crossing constraint is naturally implemented by a hard core
repulsion between the bosons which in one dimension is equivalent to
non-interacting fermions. The Pauli principle then automatically
prevents crossing. If the length $L_y$ of the polymers tends to
infinity, their {\it reduced} free energy $L_y E_0/\hbar$ is given by
ground state energy $E_0=(\pi^2/6) (\hbar^2/m) \rho^3 L_x$ of 1D
fermions at density $\rho$. Using the mapping $m \to g$ and
$\hbar \to T$, one gets for the reduced free energy density
of the polymers at fixed density
\begin{equation}
\label{line-free-energy}
f_{\rm{poly}}=\frac{\pi^{2}}{6}\frac{T}{g}\rho^3 \, .
\end{equation}
Although the scaling of this result with the physical parameters is
the same as for the free energy in the continuum limit of the dimer
model, Eq.~\eqref{dimer-free-energy}, the amplitudes do not agree.
However, since in the dimer model the number of polymers is not fixed,
we have to compare the dimer free energy with the potential of the
{\it grand} canonical ensemble of the polymers. The chemical potential
is obtained as $\mu = \partial f_{\rm poly}/\partial \rho = (\pi^2/2)
(T/g) \rho^2$, yielding the grand canonical potential density
\begin{equation}
\label{line-poptential}
j_{\rm poly}=f_{\rm poly}-\mu \rho=-\frac{\pi^{2}}{3}
\frac{T}{g}\rho^3 \, ,
\end{equation}
which is in full agreement with the continuum limit of the exact
solution of the dimer model of Eq.~\eqref{dimer-free-energy}. This
demonstrates that dimer model can be described on large length scales
as free fermions where their mass is determined in terms of the bond
weights by a random walk of a single polymer on the lattice.

It is instructive to compare the exact lattice result for the dimer
free energy of Eq.~\eqref{eq:f-hexa-exact} and the potential $j_{\rm
  poly}$ with $g_{\hexagon}$ of Eq.~\eqref{eq:hexa-g} even for larger
$b_{\hexagon}\rho_{\hexagon}$. By numerical integration of
Eq.~\eqref{eq:f-hexa-exact}, we obtain the ratio $f_{\rm dimer}/j_{\rm
  poly}$ over the entire range of possible polymer densities shown
in Fig.~\ref{fig:free-energy-hexa}. Up to approximately $1/4$ of the
maximal density, we find reasonable agreement between the lattice and
continuum results, almost independent of the anisotropy $\eta=z_1/z_2$. For
larger densities the value of $\eta$ becomes important.  There is an
optimal value of $\eta$ close to $1/3$ for which the continuum
description gives accurate results (within a few percent) even for all
densities.

\placefigure{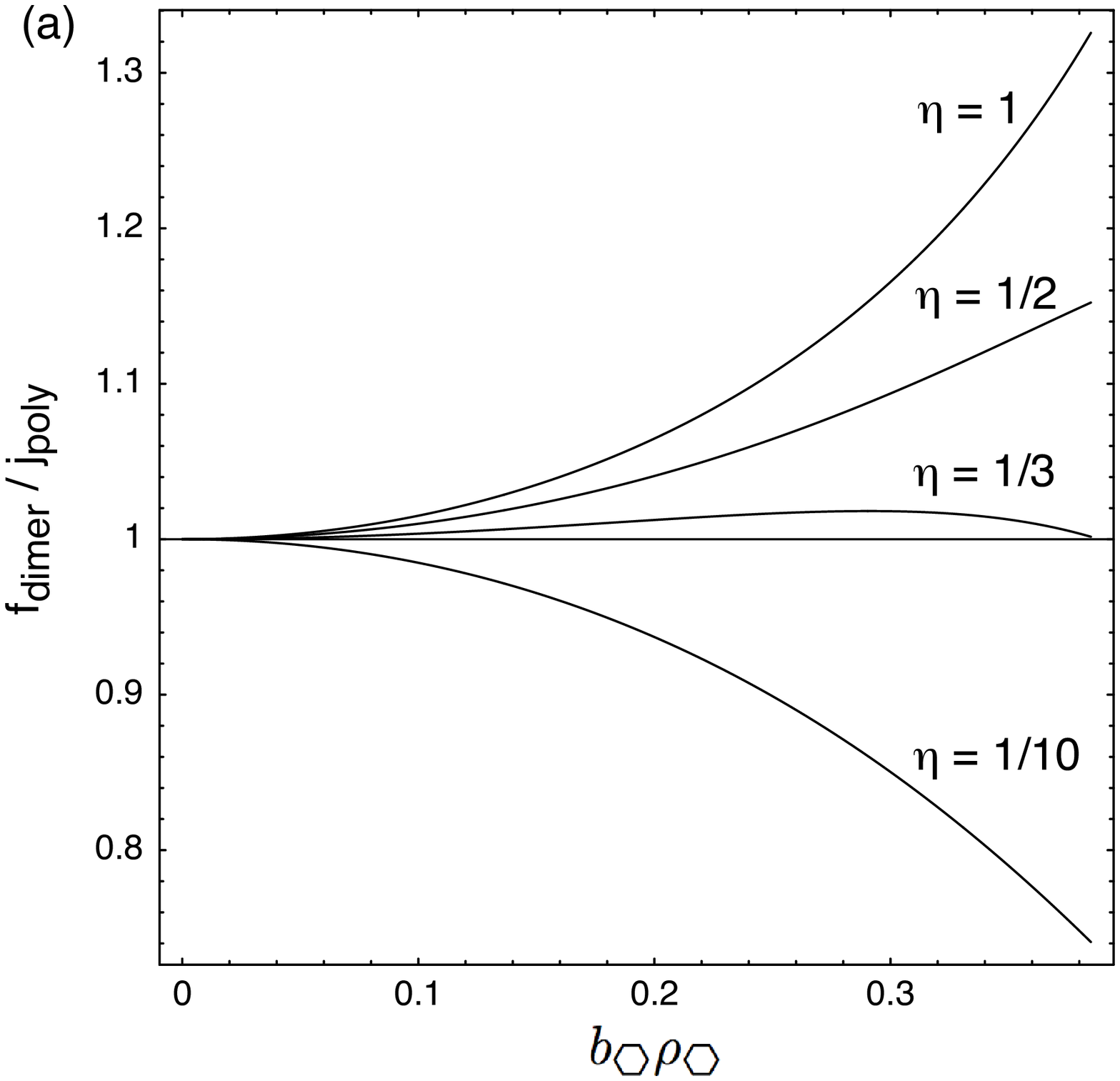}{0.85\linewidth}{Comparison of the free energy
  $f_{\rm dimer}$ of the hexagonal lattice dimer model,
  Eq.~\eqref{eq:f-hexa-exact}, and the grand canonical potential
  $j_{\rm poly}$ of the continuum polymer system,
  Eq.~\eqref{line-poptential} with the expression of
  Eq.~\eqref{eq:hexa-g} for $g/T$ substituted. The curves extend to
  $\rho_{\hexagon}=2/(3^{3/2}b_{\hexagon})$ which is the maximal
  polymer density on the lattice for $\eta=1$. However, in the lattice
  model, the density cannot become larger than $\rho_{\hexagon}=2/(\pi
  \sqrt{3})\arcsin(\sqrt{2+\eta}/2)$ by tuning the weights at a fixed
    ratio $\eta$.}{fig:free-energy-hexa}

The anisotropy of the dimer model can be tuned by changing the
relative magnitude of the weights $z_1$ and $z_2$. The exact solution
of the dimer model on the hexagonal lattice yields for the correlation
lengths the result \cite{Yokoi-Nagle-Salinas-JSP}
\begin{equation}
\label{correlation-lengths}
\xi_{x}=\frac{\sqrt{3}b_{\hexagon}}{2\phi_{0}}=\frac{1}{\pi\rho_{\hexagon}}
,\ \
\xi_{y}=\frac{3 b _{\hexagon}}{4z_{1}z_{2}\phi_{0}\sin2\phi_{0}} \,
\end{equation}
with $\phi_0=\arcsin\sqrt{((z_1+z_2)^2-1)/(4z_1z_2)}$. Hence the
correlation length perpendicular to the direction of the polymers is set
by their mean distance $1/\rho_{\hexagon}$. The length $\xi_y$ should
then be set by the typical scale a polymer can wander freely
before it reaches a transverse displacement of the order of the mean
distance between polymers.  In the continuum description of the
polymers, the random walk description of a single polymer then implies
the relation
\begin{equation}
\label{eq:aniso-single}
\xi_x^2 = \frac{T}{g_{\hexagon}} \xi_y
\end{equation}
between the correlation lengths. Together with the first
relation of Eq.~\eqref{correlation-lengths}, this yields
the anisotropy
\begin{equation}
\label{eq:aniso}
\frac{\xi_x}{\xi_y}=\pi \frac{T}{g_{\hexagon}} \rho_{\hexagon} \, .
\end{equation}
That this result is consistent with the anisotropy of the elastic
description of the polymer system follows from the macroscopic compression
modulus which is given by the compressibility, i.e.,
$c_{11}=T \rho^2 \partial^2 f_{\rm poly}/\partial \rho^2$ in terms of the
reduced free energy of Eq.~\eqref{line-free-energy}. This yields
$c_{11}=\pi^2 (T^2/g) \rho^3$ and hence together with $c_{44}=g\rho$
\begin{equation}
\label{eq:aniso-elastic}
\frac{\xi_x}{\xi_y}=\sqrt{\frac{c_{11}}{c_{44}}} \, .
\end{equation}
In order to connect the lattice and continuum description further, one would
like to know under what conditions the lattice correlations lengths of
Eq.~\eqref{correlation-lengths} fulfill the continuum relation of
Eq.~\eqref{eq:aniso-single}.  To address this question we change again
variables from $z_1$, $z_2$ to $\eta=z_1/z_2$ and
$\rho_{\hexagon}$. If one uses the result for $g_{\hexagon}$ of
Eq. \eqref{eq:hexa-g} one can easily check that
Eq.~\eqref{eq:aniso-single} is indeed fulfilled in the continuum limit
$b_{\hexagon}\to 0$. Hence, the exact anisotropy factor
$\sqrt{c_{11}/c_{44}}$ of the continuum elastic model is recovered.
This is important if one compares free energy {\it densities}
of systems with different anisotropies since then the ratio of
system sizes must be chosen as to match the ratio of their
anisotropies.

For the dimer model on the square lattice, there is no direct analog
of the previous analysis since the mean polymer density
cannot be tuned by changing the weight $z_1$ but is fixed, see
Eq.~\eqref{eq:square-density}. Hence, one cannot take the continuum
limit explicitly. Nevertheless, the square lattice is particular
useful if one wants to study the effect of random bonds since
the polymer density is robust against variations of the weights and
thus shows no disorder induced fluctuations. For the clean
square lattice dimer model, we can use the insight we gained
from the previous analysis of the hexagonal lattice to compare
the free energies of the square lattice and the continuum model.
The exact reduced free energy density of the lattice model is known to
be \cite{Kasteleyn-61-Fisher-61}
\begin{equation}
\label{eq:f-exact-square}
f_{\rm dimer}=-\frac{1}{\pi b_\Box^2} \int_0^{z_1} dv \, \frac{\arctan v}{v} \, .
\end{equation}
The continuum free fermion result of Eq.~\eqref{line-poptential}
yields in combination with the random walk result of
Eq.~\eqref{eq:square-g} for the line tension on the square lattice the
estimate for the reduced free energy density
\begin{equation}
\label{eq:f-cont-square}
f_{\rm dimer}=-\frac{\pi^2}{3} \frac{z_1 b_\Box}{z_1+1/2}\, \rho_\Box^3 \, .
\end{equation}
Below we will study the square lattice with bond energies that are
randomly distributed with mean zero so that the clean limit
corresponds to the isotropic case $z_1=1$. For this case
Eq. \eqref{eq:f-exact-square} yields the exact result $f_{\rm
  dimer}b_\Box^2=-G/\pi =-0.2916$ which is close to the continuum
approximation of Eq.~\eqref{eq:f-cont-square} which predicts $f_{\rm
  dimer}b_\Box^2=-\pi^2/36=-0.2742$ at the fixed density $\rho_{\Box}=1/(2b_{\Box})$.

%%%
%%% disordered systems
%%%
\section{Random bonds and pinned polymers}

In this section we will consider exclusively the square lattice but
with random energies $\epsilon_{ij}$ assigned to all vertical bonds so
that $z_{ij}=\exp(-\epsilon_{ij}/T_d)$, where $z_{ij}$ denotes now the
weight on the bond $(ij)$. The dimer temperature $T_d$ measures the
strength of disorder.  The energies $\epsilon_{ij}$ are drawn for each
bond independently from a Gaussian distribution with zero mean and
unit variance.  On all horizontal bonds we set $\epsilon_{ij}=0$ so
that for $T_d \to \infty$ the isotropic clean dimer model is
recovered. The partition function can be
written as
\begin{equation}
\label{eq:random-partfct}
Z_{\Box}=\sum_{\{D\}} \exp\left(-\sum_{(ij)\in D} \epsilon_{ij}/T_d \right) \, ,
\end{equation}
where the second sum runs over all occupied bonds.  The disorder
averaged free energy and correlations of this model have been computed
by a polynomial algorithm
\cite{Zeng-Leath-Hwa-PRL,Elkies-Kuperberg-Larsen-Propp-JAC} for system
sizes up to $512 \times 512$ lattice sites and typically $6000$
disorder samples \cite{Zeng-Leath-Hwa-PRL,Bogner-Emig-Taha-Zeng-PRB}.

On the analytical side, progress has been made for the polymer system
with random pinning by applying the replica method. Regarding again
each polymer of the replicated theory as a fermion in imaginary time,
and applying the Pauli principle for all particles within the same
replica, the replica free energy can be obtained again as the ground
state energy of a one-dimensional system of fermions.  Due to the
replication, the fermions carry now $n$ spin components and interact
via an attractive $\delta$-function potential arising from the short
ranged disorder correlations. This $SU(n)$ fermi gas can be studied by
a series of nested Bethe {\it Ans\"atze}
\cite{Kardar-NPB}.
In the limit $n \to 0$, the Bethe {\it Ansatz} equations can be solved
exactly for arbitrary disorder strength $\Delta$, yielding the
disordered averaged reduced free energy density of the polymers
\cite{Emig+03},
\begin{equation}
\label{eq:quenched-average-free-energy}
\bar{f}_{\rm poly}=\bar{f}_{0}(\Delta)\rho+\frac{\pi^{2}}{6}
\frac{T}{g}\rho^{3}+\frac{\Delta}{2T^2}\rho^{2} \, ,
\end{equation}
where $\bar{f}_0(\Delta)$ represents the disorder dependent free
energy of a single polymer. Notice the simple form of the disorder
contribution to the free energy of the pure system,
cf.~Eq.~\eqref{line-free-energy}.  Interestingly, in the limit of
strong interactions (disorder) the $SU(n)$ fermi gas in the limit $n
\to 0$ becomes identical to the (pure) interacting Bose gas studied by
Lieb and Liniger \cite{Lieb+Liniger}. Since it was shown that the interaction strength
scales as $n^2$, perturbation theory for the ground state energy of
the Bose gas yields a series expansion in $n$ of the replica free
energy for large disorder. The coefficients of this expansion
correspond to the disorder averaged cumulants of the free energy which
hence are known exactly from the replica Bethe {\it Ansatz}
\cite{Emig-Kardar-NPB}.

The prediction of the replica approach can be compared to the
numerical evaluation of the free energy and its cumulant averages for
the random bond dimer model. This has been done in
Ref.~\onlinecite{Bogner-Emig-Taha-Zeng-PRB}, neglecting however fluctuations
in the polymer density induced by the statistics of the pure dimer
model. While nice agreement was found for the second and third
cumulant of the free energy, the averaged free energies were only
consistent if a term $\sim\Delta^2$ of the single polymer contribution
$\bar{f}_0$ was dropped by hand. However, contributions $\sim
n^3\Delta^2$ from the replica free energy of a single polymer were
found to be crucial for the agreement of the third cumulant of the
total free energy.  A similar observation was made
\cite{Bogner-Emig-Taha-Zeng-PRB} for the data obtained previously for
a single pinned polymer \cite{Krug+92}. This is in particular
unsatisfying due to the model character of the directed polymer in a
random potential for the theory of disordered systems. Below, we show
that the differences between the replica approach for polymers and the
numerical results on the dimer model can be fully reconciled when
polymer density fluctuations are included. As demonstrated for the
clean system, this can be done by comparing the canonical dimer
ensemble to the grand canonical ensemble of polymers. From
Eq.~\eqref{eq:quenched-average-free-energy} follows the disorder
averaged chemical potential $\bar{\mu}=\partial \bar{f}_{\rm
  poly}/\partial \rho$ since disorder induces no (additional)
fluctuations in $\rho$. Thus, the average grand canonical potential
density is
\begin{equation}
\label{dimer-partition-functions}
\bar{j}_{\rm poly}
=  -\frac{\pi^{2}}{3}\frac
{T}{g}\rho^{3}-\frac{\Delta}{2T^{2}}\rho^{2}  \, ,
\end{equation}
so that the single polymer term $\bar{f}_0$ cancels.  Notice that $\bar{f}_0$
is exactly the term which was in disagreement with numerical results
for the random bond dimer model. Hence, the disorder averaged free
energy of a single directed polymer cannot be determined from
numerical computations of the free energy of the dimer model. However,
it is the disorder induced effective {\it interaction} of the polymers
which determines the dimer free energy. To obtain the latter, we
substitute the dimer parameters $\rho=\rho_\Box=1/(2b_\Box)$ and $g/T$
from Eq.~\eqref{eq:square-g} into
Eq.~\eqref{dimer-partition-functions}.  The disorder strength
$\Delta/T^2$ must be related to the dimer temperature $T_d$ which
measures the variance of the bond energies. As was shown in
Ref.~\onlinecite{Bogner-Emig-Taha-Zeng-PRB} the relation is
\begin{equation}
\frac{\Delta}{T^2}=\frac{\xi_{d}}{b_\Box}\frac{1}{T_d^2}\, ,
\label{disorder-strength}
\end{equation}
where the length $\xi_d$ acts as a cutoff in the continuum model over
which the $\delta$ function of Eq.~\eqref{eq:VV-corr} is smeared out.
The ratio
$\xi_d/b_\Box$ must be considered as a fitting parameter which should
turn out to be of order unity.  As we are comparing energy {\it
  densities}, the disorder dependent anisotropy $\sqrt{c_{11}/c_{44}}$
of the polymer system must be included \cite{Bogner-Emig-Taha-Zeng-PRB}.
This yields for the disorder averaged free energy density of
the dimers
\begin{equation}
\label{eq:f-dimer-dis}
\bar{f}_{\rm dimer}=b_\Box^2 \, \sqrt{\frac{c_{44}}{c_{11}}}
\, \bar{j}_{\rm poly} \, .
\end{equation}
The compression modulus can be obtained again from the (averaged)
polymer free energy of Eq.~(\ref{eq:quenched-average-free-energy}),
$c_{11}=T\rho^2\partial^2 \bar{f}_{\rm poly}/\partial \rho^2$,
yielding for the anisotropy
\begin{equation}
\sqrt{\frac{c_{44}}{c_{11}}}=
\frac{1}{\pi\rho}\frac{g}
{T}\left(  1+\frac{g\Delta}{\pi^{2}T^{3}\rho}\right)^{-1/2} \, .
\label{rescaling-factor}
\end{equation}
In order to correct for a (small) difference between lattice and
continuum model, we match with the exact result $f_{\rm dimer}=-G/\pi$
of Eq.~(\ref{eq:f-exact-square}) with $z_1=1$ in the clean limit
$T_d\to\infty$. Then we obtain in terms of the dimer parameters the
final result
\begin{equation}
\label{final-dimer-partition-function}
\bar{f}_{\rm dimer}=-\frac{G+\frac{3}{8}\frac{\xi_d}{b} T_d^{-2}}
{\sqrt{\pi^2+3(\xi_d/b)T_d^{-2}}} \, ,
\end{equation}
which has to be compared to the result for $\overline{\ln
  Z_l}/L^2=-\bar{f}_{\rm dimer}$ of Eq.~(32) in
Ref.~\onlinecite{Bogner-Emig-Taha-Zeng-PRB}. The expression of
Eq.~\eqref{final-dimer-partition-function} is
plotted in Fig.~\ref{fig:f-av-dimer} together with simulation data for
the random dimer model, demonstrating indeed
nice agreement for $\xi_d/b_\Box=1.33$.

\placefigure{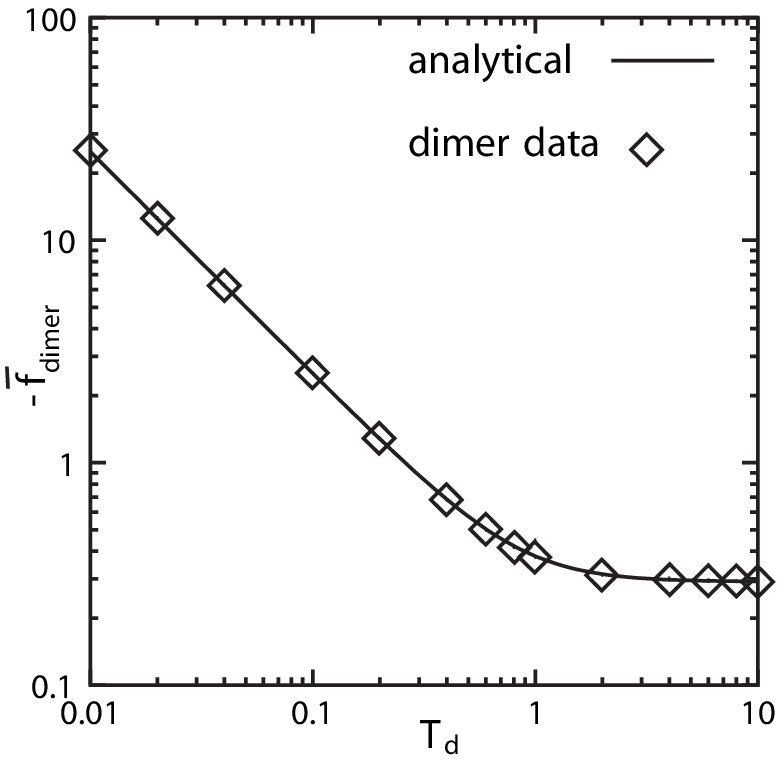}{0.8\linewidth}{ Disorder averaged free energy
  density for the random bond dimer model,
  Eq.~\eqref{final-dimer-partition-function} with $\xi_d/b_\Box=1.33$, and
  corresponding simulation data, taken from
  Ref.~\onlinecite{Bogner-Emig-Taha-Zeng-PRB}.  }{fig:f-av-dimer}

Finally, we comment on higher cumulant averages of the dimer free
energy. They were also measured in simulations, and were shown to agree
with the free energy fluctuations of the polymer system
\cite{Bogner-Emig-Taha-Zeng-PRB}. This can be easily understood from
the fact that on the square lattice there are no sample-to-sample
variations of the mean polymer density. Hence the shift of the polymer
free energy by $-\mu \rho$ in Eq.~(\ref{line-poptential}) is
independent of disorder so that one has identical disorder averaged
cumulants in the canonical and grand canonical ensembles,
$[j^p]_c=[f^p]_c$ for $p\ge 2$, where $[\ldots]_c$ denotes a cumulant
average over disorder. Because of that, {\it fluctuations} of the
single polymer free energy $f_0$ are important for the dimer model.
This explains why contributions $\sim n^3\Delta^2$ from a single
polymer to the replica free energy had to be included in
Ref.~\onlinecite{Bogner-Emig-Taha-Zeng-PRB} to obtain agreement for
the third cumulant of the free energy between polymer and dimer model.

%%%
%%% Outlook
%%%

\section{Outlook}

We have shown that a continuum polymer model can provide a good
approximation to dimer models on bipartite lattices.  Although the
polymer description yields in general not the exact result, it
provides a more physical picture of the dimer model as its exact
solution which, moreover, is available only in the clean limit.
Lattice Ising spin models with geometric frustration can be mapped
at zero temperature exactly to classical dimer models on the dual
lattice \cite{Bloete}. The effect of thermal fluctuations and/or a
transverse magnetic field can be understood in terms of
topological defects in the polymer representation of the dimer
model \cite{Jiang-Emig-preprint-2005}. Here we have shown that the
influence of random bonds in the dimer model can be described as
pinning of polymers. This implies that the glassy state of certain
spin models with random couplings could be related to the glass
phase of polymers in a random environment. For example, it can be
easily checked that random dilution of the triangular Ising
antiferromagnet leads to a pinning of the polymers at the
non-magnetic lattice sites. Another potential application of our
results is the study of classical dimers which in addition to the
hard-core repulsion interact in a more general way. Since the
mapping to polymers is independent of the dimer interaction, one
can use the analogy between directed polymers and world lines of
bosons in imaginary time to explore dimer interactions in terms of
interacting bosons in one dimension. Recently, the classical limit
(without kinetic term) of the quantum dimer model on the square
lattice has been shown to have a phase transition between a
critical and a columnar phase due to the aligning interaction
\cite{Alet+05}. The correlations in the critical phase are found
to decay with an exponent that varies continuously with the
interaction amplitude. Using the mapping to world lines of bosons,
that exponent is determined by the compressibility of the bose gas
which presumably can be modeled by a tight-binding Hamiltonian
with an infinite on-site repulsion and a nearest neighbour
interaction \cite{Affleck+04}.

\acknowledgements

YJ is indebted to D. Baeriswyl for interesting discussions.

\end{document}